# Carbon magneto-ionics: Control of magnetism through voltage-driven carbon transport

Z. Tan[1,2,3,*], Z. Ma[3,*], S. Privitera[3], M. O. Liedke[4], E. Hirschmann[4], A. Wagner[4], J. L. Costa-Krämer[5], A. Quintana[3], A. Garcia-Tort[6], J. Herrero-Martín[6], H. Tan[3], I. Fina[7], F. Sánchez[7], A. F. Lopeandia[3,8], J. Nogués,[8,9] E. Pellicer[3], J. Sort [3,9,*], E. Menéndez[3,*]

[1]Yiwu Research Institute of Fudan University, Yiwu, 322000, Zhejiang, People's Republic of China. [2]International Institute of Intelligent Nanorobots and Nanosystems, Fudan University, Shanghai 200438, People's Republic of China. [3]Departament de Física, Universitat Autònoma de Barcelona, E-08193 Cerdanyola del Vallès, Spain. [4]Institute of Radiation Physics, Helmholtz-Zentrum Dresden – Rossendorf, Dresden 01328, Germany. [5]IMN-Instituto de Micro y Nanotecnología (CNM-CSIC), Isaac Newton 8, PTM, 28760 Tres Cantos, Madrid, Spain. [6]ALBA Synchrotron Light Source, 08290 Cerdanyola del Vallès, Spain. [7]Institut de Ciencia de Materials de Barcelona, ICMAB-CSIC, Campus UAB, 08193 Bellaterra, Barcelona, Spain. [8]Catalan Institute of Nanoscience and Nanotechnology (ICN2), CSIC and BIST, 08193 Barcelona, Spain. [9]Institució Catalana de Recerca i Estudis Avançats (ICREA), Pg. Lluís Companys 23, E-08010 Barcelona, Spain.

*e-mail: tanzw@ywfudan.cn (Z. Tan), ma.zheng@uab.cat (Z. Ma), jordi.sort@uab.cat (J. Sort), enric.menendez@uab.cat (E. Menéndez)

Control of magnetism through voltage-driven ionic processes (*i.e.*, magneto-ionics) holds potential for next-generation memories and computing. This stems from its non-volatility, flexibility in adjusting the magnitude and speed of magnetic modulation, and energy efficiency. Since magneto-ionics depends on factors like ionic radius and electronegativity, identifying alternative mobile ions is crucial to embrace new phenomena and applications. Here, the feasibility of C as a prospective magneto-ionic ion is investigated in a Fe-C system by electrolyte gating. In contrast to most magneto-ionic systems, Fe-C presents a dual-ion mechanism: Fe and C act as cation and anion, respectively, moving uniformly in opposite directions under an applied electric field. This leads to a 7-fold increase in saturation magnetization with magneto-ionic rates larger than 1 emu·cm$^{−3}$·s$^{−1}$, and a 25-fold increase in coercivity. Since carbides exhibit minimal cytotoxicity, this introduces a biocompatible dimension to magneto-ionics, paving the way for the convergence of spintronics and biotechnology.

Magneto-ionics, which deals with the control of magnetic properties through voltage-driven ionic transport and/or redox processes, is gaining momentum as a magneto-electric mechanism for low-power spintronics[1]. The appeal lies in its non-volatility, enabled by its electrochemical character[2], and its versatility in tuning the degree and speed of magnetic modulation. This is attractive for a broad range of applications, such as neuromorphic computing[3], and data encryption[4]. The diverse responses exhibited by magneto-ionic systems depend on the mobile species (*e.g.*, $H^+$[5], $Li^+$[6], $O^{2-}$[7,8], $F^-$[9], $OH^-$[10], or $N^{3-}$[11-13]), the phase and/or stoichiometry of the actuated material, and the magnetic property (*e.g.*, magnetic anisotropy[5,7], coercivity, $H_C$[14], or saturation magnetization, $M_S$[8,11,12]) used to track the magneto-ionic response. Despite significant progress in this area, the number of materials exhibiting magneto-ionics remains limited. Since magneto-ionics critically depends on the synergy of different parameters, such as the ionic radius, electronegativity, or diffusivity, the search for additional magneto-ionic materials and mobile ion species is essential to enable new phenomena and innovative applications.

Here, we demonstrate simultaneous voltage-driven C and Fe ion transport as an approach to modulate ferromagnetism in Fe-C-based heterostructures. This transport occurs uniformly, resembling a planar front, with C and Fe moving in opposite directions, behaving as cation and anion, respectively, due to their electronegativity difference. Voltage actuation results in a 7-fold and a 25-fold increase in $M_S$ and $H_C$, respectively, with a magneto-ionic rate larger than 1 emu·cm$^{-3}$·s$^{-1}$. This combination of features sets the stage for innovative uses of magneto-ionic systems and highlights the importance of investigating alternative mobile cations and anions to broaden the functionalities and applications of magneto-ionics.

The as-prepared Fe-C film comprises a heterostructure which consists of four Fe-C layers of different compositions capped by a Ti-C layer, as shown schematically in Fig. 1a (see Supplementary Section 1, Fig. S1, and Fig. S2 for details on the fabrication of the as-prepared heterostructure and the voltage actuation protocol, respectively). The as-prepared film is weakly magnetic, with a $M_S$ of 169 emu·cm$^{-3}$ and a $H_C$ of 11 Oe (Fig. 1b). Additionally, the as-prepared films exhibit moderate resistivity (Supplementary Section 2 and Fig. S3), which makes the system optimal for voltage actuation through magneto-ionics[12]. Fig. 1c shows the evolution of the change in magnetic moment at saturation (see Methods, Fig. S4, and Supplementary Section 3 for the justification of the use of magnetic moment), $\Delta m_S(t)$, while applying a gating voltage of −50 V. At the early stages of voltage actuation, the generated magnetic moment increases rapidly and linearly, evidencing strong magneto-electric effects which could be compatible with C and/or Fe ionic transport. Considering the layered structure of the as-prepared film, the electric field actuation perpendicular to the film, and the initial linear dependence of the generated magnetic moment at saturation with time (*i.e.*, $\Delta m_S(t) \propto t$), interface-controlled ionic transport is envisaged[15]. With further gating time, this growth levels off until reaching a steady state. To shed more light on ion mobility mechanisms, the data in Fig. 1c was fitted to the following Avrami formalism[15]: $\Delta m_S(t) = \Delta m_S^{t\to\infty}(1 - e^{-k(t-t_0)^n})$, where $\Delta m_S^{t\to\infty}$ is the generated long-term magnetic moment at saturation ($\Delta m_S^{t\to\infty} = (5.170 \pm 0.006)\ 10^{-4}$ emu), $k$ an overall rate constant ($k = 0.215 \pm 0.003$), $n$ the Avrami exponent ($n = 0.801 \pm 0.006$), and $t_0$ the time in which the voltage was set ON ($t_0 = 8.54 \pm 0.02$ min). An $n$ lower than one is typically associated with heterogenous nucleation and restricted growth processes[15]. However, the fit, with an adjusted R² of 0.9981, does not completely align with the experimental data, indicating that there may be different ion motion mechanisms over the extended voltage actuation periods. As shown in Fig. S5, when the Avrami formalism was applied to the generated magnetic moment data from the first 13 minutes of voltage actuation, the goodness-of-fit, with R² of 0.9998, improves considerably. An Avrami exponent of $n = 1.296 \pm 0.008$ is determined, indicating the presence of a distinct ion motion mechanism in the early stages of the magneto-ionic process, which goes beyond one-dimensional motion, suggesting a two-dimensional motion (plate-like growth)[15], as seen in nitride-based systems[11]. To estimate the magneto-ionic rate, which accounts for the change in generated magnetization with time, in the first stages of the gating, the initial slope of Fig. 1c curve is normalized to the volume of the as-prepared sample, resulting in values larger than 1 emu·cm⁻³·s⁻¹. This value is comparable to the performance of nitrogen magneto-ionics and surpasses that of oxygen magneto-ionics[8]. After applying a negative bias of −50 V for 1 h, the voltage was turned off, and a hysteresis loop was recorded by vibrating sample magnetometry (Fig. 1d). The ascending branch of the loop shows a slightly lower $M_S$ (around 2 %) compared to the descending branch, which could be compatible with a partial magnetization depletion due to a mild recarburization process. Despite this, the voltage-induced ferromagnetic changes are significant and largely permanent. Upon voltage actuation, $M_S$ and $H_C$ increase from 169 to 1176 emu·cm⁻³ and from 11 to 294 Oe, respectively. Given that Fe₃C has an $M_S$ of 960 emu·cm⁻³ [16,17], significantly lower than that of α-Fe (about 1710 emu·cm⁻³)[18], the $M_S$ of the as-prepared sample (169 emu·cm⁻³) is consistent with the formation of paramagnetic Fe₂C and ferromagnetic Fe₃C (Fig. 1b). In contrast, as seen in Fig. 1d, the increased saturation magnetization upon voltage actuation (with a value of 1176 emu·cm⁻³, a 7-fold increase compared to the value of the as-prepared sample) is consistent with the formation of metallic iron and Fe-rich phases, aligning with voltage-driven Fe and/or C ion transport processes that decarburize certain regions of the film. Furthermore, the increase in coercivity from 11 to 294 Oe after voltage actuation (Fig. 1d) is compatible with the formation of ferromagnetic Fe clusters[19] surrounded by Fe carbides, since $H_C$ is proportional to the volume of magnetic material above the superparamagnetic limit and below the critical single-domain size[18,20], and more stoichiometric Fe₃C[21].

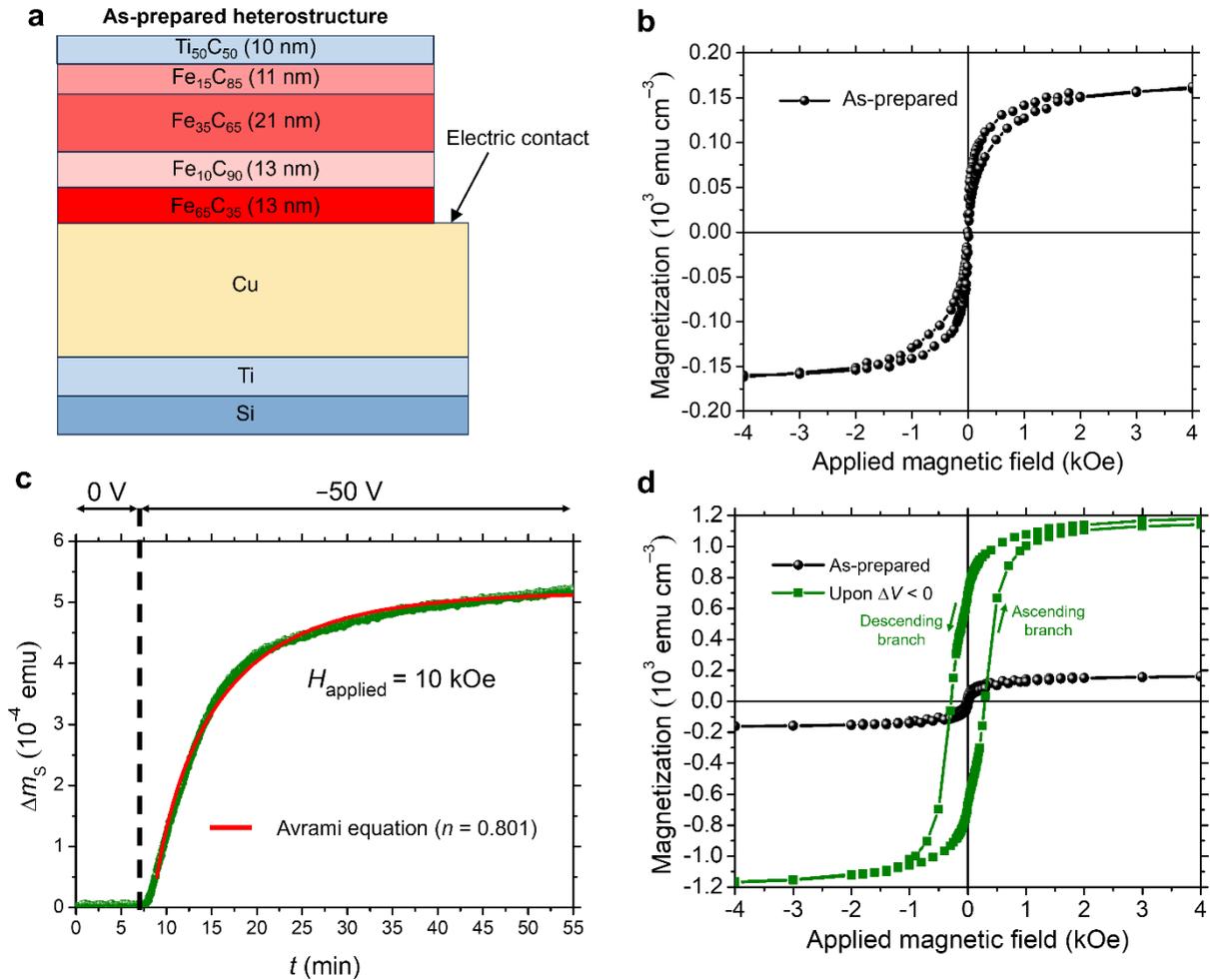

**Fig. 1 | Magneto-ionic response of the as-prepared heterostructure. a**, Schematic illustration of the cross-section of the as-prepared heterostructure. **b**, Hysteresis loop of an as-prepared heterostructure. **c**, Dependence of the generated magnetic moment at saturation (*i.e.*, $\Delta m_S$) as a function of time ($t$) while applying –50 V. The experimental datapoints are fitted to an Avrami equation. **d**, Hysteresis loops of the as-prepared sample (shown in panel **b**), and after being actuated with –50 V for 1 h. The magnetic field was swept from 20 kOe to −20 kOe and back, and the arrows indicate descending and ascending branches. The magnetization in panels **b** and **d** is normalized to the volume of the Fe-containing layers (see Methods).

To validate the hypothesis of voltage-driven ion motion of Fe and/or C in a planar-like front, a high-angle annular dark-field scanning transmission electron microscopy (HAADF-STEM)/elemental electron energy loss spectroscopy (EELS) characterization of an as-prepared sample and a sample treated with −50 V for 1h was carried out (Fig. 2). Energy-dispersive X-ray spectroscopy (EDX) characterization shows a four-layer structure in the as-prepared state, consisting of $Fe_{15}C_{85}$ (Area I), $Fe_{35}C_{65}$ (Area II), $Fe_{10}C_{90}$ (Area III), and $Fe_{65}C_{35}$ (Area IV) layers from top to bottom. When voltage is applied, the original multilayer structure (with four distinct Fe-C layers) transforms into a bilayer, where the bottom layer, which resembles the bottom layer of the as-prepared sample from compositional and microstructural viewpoints, becomes thicker (Fig. S6 and Fig. S7 for further information on microstructural aspects). When comparing the composition of the as-prepared and −50 V-gated samples, a well-defined trend of ion diffusion is manifested. The voltage-induced top layer (Area V) is virtually Fe-free and is of the same thickness as the sum of thickness of Area I, Area II, and Area III of the as-prepared sample, while the voltage-induced bottom layer (Area VI) becomes thicker than Area IV with a richer composition in Fe: $Fe_{80}C_{20}$. These results clearly evidence that the voltage induces dual-ion motion, where Fe and C move in opposite directions. Namely, C, being more electronegative than Fe, moves upwards (towards the counter electrode which is positively polarized with respect to the Cu bottom electrode, Fig. S2), whereas Fe migrates downwards. Moreover, after biasing, the Ti-C layer becomes richer in C (*i.e.*, $Ti_{50}C_{50}$ *vs*. $Ti_{40}C_{60}$ for the as-prepared and voltage-treated samples, respectively), further confirming the voltage-driven C transport towards the upper part (*i.e.*, to the Ti-C reservoir).

To unravel the crystalline order, high-resolution TEM imaging was carried out (Fig. S6 and Fig. S7). The C-rich layers (*i.e.*, Area I, Area II, Area III —as-prepared sample—, and Area V —voltage-treated sample—) are amorphous-like (see Fig. S6 in which a diffuse halo and no rings nor spots are observed in the fast Fourier transform, FFT, images), while the others (*i.e.*, Area IV —as-prepared— and Area VI —voltage-treated—) exhibit some degree of crystallinity (Fig. S7). FFTs reveal Fe carbide formation in the form of $Fe_2C$ and $Fe_3C$ in Area IV of the as-prepared heterostructure. The interplanar distances are slightly shifted towards slightly lower values, confirming the formation of non-stoichiometric Fe carbides, as envisaged from the magnetometry results. With voltage, slightly more polycrystalline $Fe_2C$ and $Fe_3C$ phases form, since more diffraction spots appear, and spot #3 in Fig. S7 is likely arising mainly from metallic Fe, revealing the effect of Fe and C transport in the microstructure (Fig. S7). The presence of oxide traces is mainly linked to natural oxidation of the prepared lamellae.

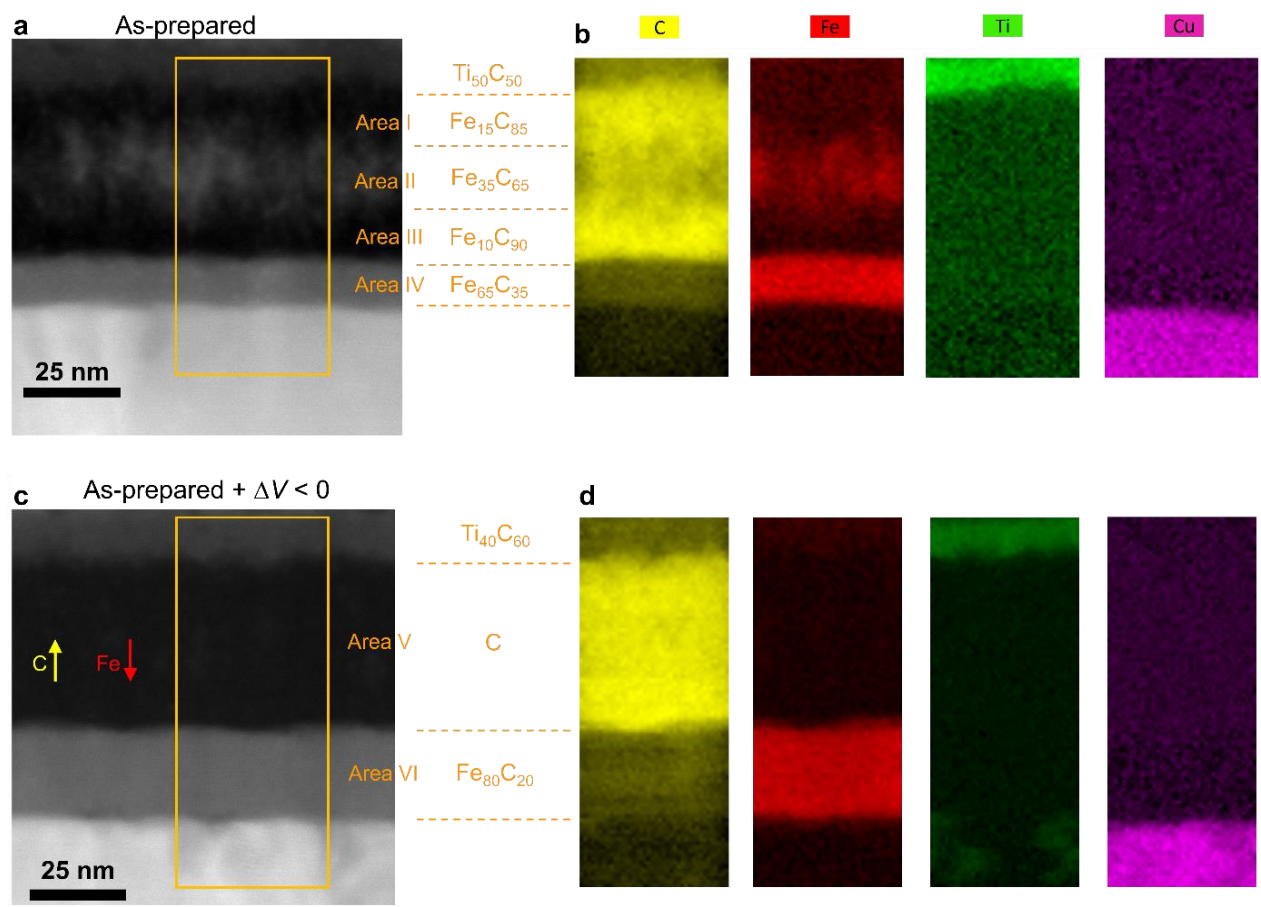

**Fig. 2 | Compositional characterization of as-annealed and voltage-actuated heterostructures. a**, HAADF-STEM micrograph of a cross section of an as-prepared sample. **b**, C, Fe, Ti, and Cu elemental EELS mappings of the region marked with a yellow rectangle in panel a. **c**, HAADF-STEM micrograph of a cross section of a sample treated with –50 V for 1 h. **d**, C, Fe, Ti, and Cu elemental EELS mappings of the region marked with a yellow rectangle in panel c.

The evolution of defect microstructure as a function of depth was determined by variable energy positron annihilation lifetime spectroscopy (VEPALS). As seen in Fig. 3, the main changes in the VEPALS occur in the bottom layer of the gated samples (Area VI). The average positron lifetime clearly decreases upon voltage actuation, indicating smaller average defect sizes, in concordance with the enhanced crystallinity and increased thickness of Area VI (23 nm) compared to Area IV (13 nm), (see Fig. S4, Fig. S6, and Fig. S7). The estimated major defect size (*i.e.*, $\tau_1$) reduces from the range of bi-vacancy for as-prepared samples to the single vacancy in case of the biased sample (Fig. 3b). In addition, vacancy agglomerations (*i.e.*, $\tau_2$), originating from the amorphous-like counterparts, are found, undergoing a strong reduction in size from ≈ 16 to ≈ 12 vacancies within the vacancy complex upon voltage actuation (Fig. 3b). See Methods and Supplementary Section 4 for further details on the VEPALS characterization.

Remarkably, the HR-TEM and VEPALS results seem to indicate not only an ion movement, but an unprecedented partial voltage-induced recrystallization of some of the amorphous layers in the system.

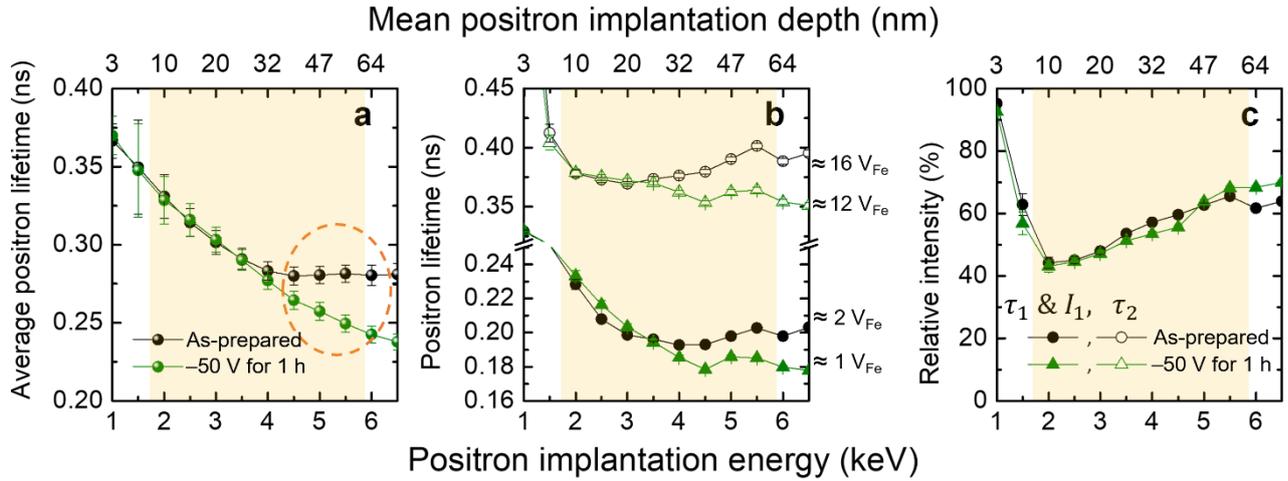

**Fig. 3 | Defect characterization by VEPALS. a,** Average positron lifetime. **b,** Positron lifetime components (*i.e.*, $\tau_1$ and $\tau_2$), and **c** the relative intensities of defects characterized by $\tau_1$ (*i.e.*, $I_1$) as a function of positron implantation energy (*i.e.*, mean positron implantation depth, see Methods) of the as-prepared and voltage-treated heterostructures. The orange background approximates the Fe-C film region (see Methods), and the orange dashed-line mainly embraces the results corresponding to Area IV of the as-prepared sample and Area VI of the voltage-treated one. $V_{Fe}$ stands for iron vacancy.

To gain more insight into the compositional aspects, X-ray absorption spectroscopy (XAS) measurements were performed, in both total electron yield (TEY) and fluorescence yield (FY) modes (Fig. S8). Total electron yield (TEY) characterizes the oxidation state of the first top nanometers, while fluorescence yield (FY) provides insights into much deeper parts (*i.e.*, several tens of nm)[22]. Fig. S8a presents the Fe $L_{2,3}$-edge XAS spectra for the as-prepared, and voltage-treated with −50 V for 5 min heterostructures. The as-prepared sample spectra reveal characteristic peaks at around 707.3 and 708.8 eV at the $L_3$ edge, compatible with $Fe^{2+}$ and $Fe^{3+}$ which are linked to iron carbide formation, in agreement with the high-resolution TEM results (Fig. S6). Upon voltage actuation, the spectra undergo noticeable changes, as highlighted in Fig. S8b. Interestingly, it can be seen that $L_3/L_2$ ratio, decreases in TEY mode, whereas it increases in FY mode. Since the $L_3/L_2$ ratio is inversely proportional to the degree of carburization (*i.e.*, to the C/Fe ratio)[23], this indicates an increased C/Fe ratio at the film surface with respect to the rest of the heterostructure, confirming the voltage-driven dual motion of C (upwards) and Fe (downwards).

Summarizing, our findings unambiguously demonstrate a planar-front voltage-driven dual-ion (C and Fe) transport as a means to control ferromagnetism in Fe-C-based heterostructures. This dual-ion movement leads to a large increase in saturation magnetization and coercivity with rather large magneto-ionic rates. Notably, since carbon, iron, and their carbides exhibit low cytotoxicity, this opens new possibilities for the integration of spintronics and biotechnology[24], such as brain-machine interfaces[25], offering a communication bridge between the brain's electrical activity and an external device.

## Methods

### Sample fabrication: growth of the heterostructure

45 nm-thick Fe-C thin films were grown at room temperature using DC and RF co-sputtering from Fe and C targets, respectively, on top of Si(001) wafers previously coated with a 15 nm-thick Ti adhesion layer and a 50 nm-thick Cu seed layer (to act as bottom electrode). Before depositing the Fe-C film, a portion of the Cu seed layer was masked to make electrical contacts for subsequent magneto-electric characterization (Fig. 1a). A protective 5 nm-thick layer of Ti-C was deposited on the Fe-C layer by co-sputtering of Ti and C targets using DC and RF powering, respectively. The substrate-to-target distance was approximately 10 cm. The base pressure of the system was around $8 \times 10^{-8}$ Torr. The depositions were carried out in an AJA International, Inc. ATC 2000-V Sputtering System. Subsequently, the films were annealed in a tubular furnace at 300 °C for 2 h under a controlled atmosphere of 95 % Ar and 5 % $H_2$ at a total flow rate of 80 sccm (flow percentages). The heating and cooling rates were 3 °C/min. See Supplementary Section 1 and Fig. S1 for further details on the fabrication of the as-prepared heterostructure. The as-grown samples after annealing are denoted as "as-prepared" throughout the manuscript.

### Magneto-electric characterization

Magnetic measurements were carried out using a vibrating sample magnetometer (VSM) from MicroSense (LOT, Quantum Design). Magneto-electric measurements were done by performing VSM measurements while applying voltage through a custom-made electrolytic cell (Fig. S2). Electrolyte gating of the samples was achieved using an external Agilent B2902A power supply, applying voltage across the counter electrode (a Pt wire) and the working electrode (*i.e.*, the Cu bottom electrode). The electrolyte utilized was anhydrous propylene carbonate solvating $Na^+$ and $OH^-$ species (10 - 25 ppm), generated through the immersion of metallic sodium pieces capable of reacting with any residual water traces[8,11,12]. Since both the fabrication of the as-prepared sample and the voltage treatment result in an increased thickness (Fig. S4 and Supplementary Section 3), the magnetization values are determined by normalizing the magnetic moment to the volume of Fe-containing layers, and the magnetic moment at saturation is plotted instead of saturation magnetization in Fig. 1c. All samples are of the same area (0.3 $cm^2$). The thickness of as-grown, as-prepared, and voltage-treated samples are 45, 58, and 68 nm, respectively (Fig. S4), while the thicknesses of Fe containing layers of these heterostructures are 45, 58, and 23 nm, respectively (thickness of Area VI since Area V is virtually Fe-free, Fig. S4 and Fig. S7). Hence, the volume of Fe-containing layers is $13.5 \times 10^{-7}$, $17.4 \times 10^{-7}$, and $6.9 \times 10^{-7}$ $cm^3$, respectively. The saturation magnetization values were taken at an applied magnetic field of 10 kOe, which is sufficiently high to ensure full saturation of the involved magnetic phases. For the time dependent measurements, the field was fixed at 10 kOe. The maximum applied fields for the hysteresis loops were 20 kOe. Linear slopes in the hysteresis loops at applied magnetic fields far above the saturation field of the involved ferromagnetic phases were subtracted to remove the diamagnetic and paramagnetic contributions of the non-ferromagnetic contributions, *e.g.*, substrate, buffer and capping layers, and paramagnetic Fe-C phases. All the magnetic fields were applied along the film plane direction (*i.e.*, in-plane measurements).

**Structural and compositional characterizations**

Cross-sectional lamellae, obtained by focused ion beam, of the investigated heterostructures were characterized by transmission electron microscopy (TEM). Specifically, TEM, high-resolution TEM, high-angle annular dark-field scanning transmission electron microscopy (HAADF-STEM), and electron energy loss spectroscopy (EELS) of Fig. 2, Fig. S1a,b,c, and Fig. S4 were carried out on a TECNAI F20 HRTEM/STEM microscope operated at 200 kV. High-resolution TEM of Fig. S6 and Fig. S7, and energy-dispersive X-ray (EDX) analysis (not shown) was performed on a Spectra 300 (S)TEM microscope operated at 200 kV in the Joint Electron Microscopy Center at ALBA Synchrotron. For the cross-sectional lamellae preparation by focused ion beam, an electrically conducting Pt-C is deposited on top of the heterostructures to enhance electrical conductivity (and to minimize charging effects), to protect them from oxidation and contamination, and to improve the quality of the final thin section since it prevents ion beam damage. Note that the quantification of Ti-C layer composition using EDX may be influenced by the carbon present in the adjacent Pt-C layer used for the lamella preparation. While the observed trend in carbon concentration changes within the Ti-C layers during annealing and voltage actuation is reliable, the absolute values should be interpreted as approximate values. Moreover, given that the composition quantification by EDX was performed over a relatively small area (approximately 400 nm²), the reported compositions should be considered approximate yet representative. A similar limitation applies to the color intensity in the elemental EELS mappings, where only simultaneously recorded mapping images should be compared.

X-ray absorption spectroscopy (XAS) at the Fe $L_{2,3}$ edges was carried out at the BL29-BOREAS beamline at the ALBA Synchrotron[26]. The spectra were obtained using both total electron yield (TEY) and fluorescence yield (FY) modes at room temperature (300 K).

Defect characterization was carried out by variable energy positron annihilation lifetime spectroscopy (VEPALS). VEPALS measurements were conducted at the Mono-energetic Positron Source (MePS) beamline at Helmholtz-Zentrum Dresden – Rossendorf (Germany)[27]. A CeBr$_3$ scintillator detector together with a Hamamatsu R13089-100 photomultiplier tube for the gamma photons detection was employed. A Teledyne SPDevices ADQ14DC-2X digitizer with a 14-bit vertical resolution and 2GS/s (gigasamples per second) horizontal resolution was utilized for the processing of signals. The overall time resolution of the measurement system is ≈ 0.250 ns and all spectra contain at least 1×10⁷ counts. A typical lifetime spectrum $N(t)$, which is the absolute value of the time derivative of the positron decay spectrum, is described by $N(t) = R(t) * \sum_{i=1}^{k+1} \frac{I_i}{\tau_i} e^{-t/\tau_i} + \text{Background}$, where k is the number of different defect types contributing to the positron trapping, which are related to k + 1 components in the spectra with individual lifetimes $\tau_i$ and intensities $I_i$ ($\sum I_i = 1$). The instrument resolution function $R(t)$ is a sum of two Gaussian functions with distinct intensities and relative shifts both depending on the positron implantation energy, $E_P$. It was determined by measuring a reference sample, *i.e.* yttria-stabilized zirconia, which exhibits a known single lifetime component of 182 ± 3 ps. The background was negligible. All the spectra were deconvoluted using a non-linear least-squares fitting method, minimized by the Levenberg-Marquardt algorithm in the software package PALSfit[28], into 2 major lifetime components, which directly evidence localized annihilation at 2 different defect types (sizes; $\tau_1$ and $\tau_2$). The corresponding relative intensities largely reflect the concentration of each defect type (size) if the size of compared defects is in a similar range. In general, positron lifetime is directly proportional to defects size, *i.e.*, the larger the open volume is, the lower probability there is for positrons to be annihilated with electrons, and thereby the longer positron lifetime[29]. The positron lifetime and its intensity has been probed as a function of positron implantation energy $E_P$ which can be recalculated to the mean positron implantation depth $\langle z \rangle$ using: $\langle z \rangle \text{(nm)} = \frac{23.9}{\rho\left(\frac{g}{cm^3}\right)} E_P(\text{keV})^{1.69}$[30], which is an approximate measurement of depth since it does not account for positron diffusion. The orange background of Fig. 3, which approximates the Fe-C film region, is calculated using this formula and assuming the density of $Fe_3C$[16] for the Fe-C film. The average positron lifetime $\tau_{Av}$ is defined as $\tau_{Av} = \sum \tau_i I_i$, and it has a high sensitivity to the defect size (type).

**Electrical characterization**

For the electrical actuation across the as-prepared heterostructures, voltages were applied through a TFAnalyser3000 platform (aixACCT Systems GmbH) between the bottom electrode and a probe tip that was in contact with the sample surface. The presented current intensity *vs.* applied voltage curves were obtained from averaging the datasets recorded on ten different areas of the surface of the heterostructures using 2 s integration time.

**Additional information**

Correspondence and requests for materials should be addressed to E.M.

**Author contributions**

E.M. had the original idea and led the investigation. Z.T. and E.M. designed the experiments. J.S. and E.M. supervised the work. Z.T., Z.M., and E.P. synthesized the Fe-C films. Z.T., Z.M., S.P., and A.Q. carried out the magneto-electric measurements and analyzed the data. A.F.L., J.N., J.S., and E.M. applied the Avrami formalism to the time-dependent magneto-electric characterization. Z.T., Z.M., A.G.-T., and J.H.-M. carried out the XAS characterization and analyzed the data. Z.T., J.L.C.-K., and E.P. performed the TEM and STEM characterization and carried out the analysis of the corresponding data. Z. M., H.T., I.F., and F.S. performed the out-of-plane electrical measurements and analyzed the data. M.O.L., E.H., and A.W. characterized the samples by PALS and analyzed the data. All authors discussed the results and commented on the article. The article was written by Z.T., J.N., and E.M. with contributions from all co-authors.

**Competing financial interests**

The authors declare no competing financial interests.

**Acknowledgements**

Financial support by the European Union's Horizon 2020 Research and Innovation Programme (BeMAGIC European Training Network, ETN/ITN Marie Skłodowska–Curie grant № 861145), the DOCFAM-PLUS within HORIZON-MSCA-2021-COFUND-01 (grant № 101081337), the European Research Council (2021-ERC-Advanced REMINDS Grant № 101054687), the Spanish Government (PID2020-116844RB-C21 and TED2021-130453B-C21), the Generalitat de Catalunya (2021-SGR-00651 and 2021-SGR-00804), the MCIN/AEI/10.13039/501100011033 & European Union NextGenerationEU/PRTR (grant № CNS2022-135230), and the Zhejiang Provincial Postdoctoral Research Project First-Class Funding, China, (Grant No. ZJ2024029) is acknowledged. ICN2 is funded by the CERCA program/Generalitat de Catalunya. The ICN2 is supported by the Severo Ochoa Centres of Excellence program, Grant CEX2021-001214-S, grant funded by MCIU/AEI/10.13039/501100011033. Authors acknowledge the use of instrumentation financed through Grant IU16-014206 (METCAM-FIB) to ICN2 funded by the European Union through the European Regional Development Fund (ERDF), with the support of the Ministry of Research and Universities, Generalitat de Catalunya. The XAS experiments were performed at BL29-BOREAS beamline at ALBA Synchrotron with the collaboration of ALBA staff under the in-house research proposal 2023027292. VEPALS were carried out at ELBE from the Helmholtz-Zentrum Dresden – Rossendorf e. V., a member of the Helmholtz Association. We would like to thank the facility staff for their assistance. E. M. is a Serra Húnter Fellow.

**References**

1. Nichterwitz, M. *et al*. Advances in magneto-ionic materials and perspectives for their application. *APL Mater*. **9**, 030903 (2021).

2. Leighton, C., Birol, T. & Walter, J. What controls electrostatic vs electrochemical response in electrolyte-gated materials? A perspective on critical materials factors. *APL Mater.* **10**, 040901 (2022).

3. Mishra, R., Kumar, D. & Yang, H. Oxygen-migration-based spintronic device emulating a biological synapse. *Phys. Rev. Appl.* **11**, 054065 (2019).

4. Ye, X. *et al.* Selective dual-ion modulation in solid-state magnetoelectric heterojunctions for in-memory encryption. *Small* **19**, 448–452 (2023).

5. Tan, A. J. *et al*. Magneto-ionic control of magnetism using a solid-state proton pump. *Nat. Mater.* **18**, 35–41 (2019).

6. Ameziane, M. *et al.* Lithium-ion battery technology for voltage control of perpendicular magnetization. *Adv. Funct. Mater.* **29**, 2113118 (2022).

7. Bauer, U. *et al.* Magneto-ionic control of interfacial magnetism. *Nat. Mater.* **14**, 174–181 (2015).

8. Quintana, A. *et al.* Voltage-controlled ON-OFF ferromagnetism at room temperature in a single metal oxide film. *ACS Nano* **12**, 10291–10300 (2018).

9. Vasala, S. *et al*. Reversible tuning of magnetization in a ferromagnetic Ruddlesden–Popper-type manganite by electrochemical fluoride-ion intercalation. *Adv. Electron. Mater.* **6**, 1900974 (2020).

10. Quintana, A. *et al*. Hydroxide-based magneto-ionics: electric-field control of a reversible paramagnetic-to-ferromagnetic switch in α-Co(OH)$_2$ films. *J. Mater. Chem. C* **10**, 17145–17153 (2022).

11. de Rojas, J. *et al.* Voltage-driven motion of nitrogen ions: a new paradigm for magneto-ionics. *Nat. Commun.* **11**, 5871 (2020).

12. de Rojas, J. *et al.* Critical role of electrical resistivity in magnetoionics. *Phys. Rev. Appl.* **16**, 034042 (2021).

13. Jensen, C. J. *et al*. Nitrogen-Based Magneto-ionic Manipulation of Exchange Bias in CoFe/MnN Heterostructures. *ACS Nano* **17**, 6745–6753 (2023).

14. Jeong, J., Park, Y. S., Kang, M.-G. & Park, B.-G. Nanosecond magneto-ionic control of magnetism using a resistive switching HfO$_2$ gate oxide. *Adv. Electron. Mater.* **10**, 2400535 (2024).

15. Shirzad K. & Viney C. A critical review on applications of the Avrami equation beyond materials science. *J. R. Soc. Interface* **20**, 20230242 (2023).

16. Bhadeshia, H. K. D. H. Cementite. *Inter. Mater. Rev.* **65**, 1–27 (2020).

17. Tsuzuki, A., Sago, S., Lu, J., Hirano, S.-I. & Naka, S. High temperature and pressure preparation and properties of iron carbides $Fe_7C_3$ and $Fe_3C$. *J. Mater. Sci.* **19**, 2513–2518 (1984).

18. Skomski, R. & Coey, J. M. D. Permanent Magnetism (Institute of Physics Publishing, 1999).

19. Kneller, E. F. & Luborsky, F. E. Particle size dependence of coercivity and remanence of single domain particles. *J. Appl. Phys.* **34**, 656–658 (1963).

20. Hadjipanayis, G. C. Nanophase hard magnets. *J. Magn. Magn. Mater.* **200**, 373–391 (1999).

21. Liu, D. *et al*. Giant magnetic coercivity in $Fe_3C$-filled carbon nanotubes. *RSC Adv.* **8**, 13820 (2018).

22. Sakamaki, M. & Amemiya, K. Nanometer-resolution depth-resolved measurement of florescence-yield soft x-ray absorption spectroscopy for FeCo thin film. *Rev. Sci. Instrum.* **88**, 083901 (2017).

23. Furlan, A., Jansson, U., Lu, J., Hultman, L. & Magnuson, M. Structure and bonding in amorphous iron carbide thin films. *J. Phys.: Condens. Matter.* **27**, 045002 (2015).

24. Kim, S.-B., Kim, C.-H., Lee S.-Y. & Park S.-J. Carbon materials and their metal composites for biomedical applications: a short review. *Nanoscale* **16**, 16313–16328 (2024).

25. Li, J. *et al*. Sensing and stimulation applications of carbon nanomaterials in implantable brain-computer interface. *Int. J. Mol. Sci.* **24**, 5182 (2023).

26. Barla, A. *et al*. Design and performance of BOREAS, the beamline for resonant X-ray absorption and scattering experiments at the ALBA synchrotron light source. *J. Synchrotron Rad.* **23**, 1507–1517 (2016).

27. Wagner, A., Butterling, M., Liedke, M. O., Potzger, K. & Krause-Rehberg, R. Positron annihilation lifetime and Doppler broadening spectroscopy at the ELBE facility. *AIP Conf. Proc.* **1970**, 040003 (2018).

28. Olsen, J. V., Kirkegaard, P., Pedersen, N. J. & Eldrup, M. PALSfit: A new program for the evaluation of positron lifetime spectra. *Phys. Stat. Sol. (c)* **4**, 4004–4006 (2007).

29. Tuomisto, F.& Makkonen, I. Defect identification in semiconductors with positron annihilation: experiment and theory. *Rev. Mod. Phys.* **85**, 1583–1631 (2013).

30. Dryzek, J. & Horodek, P. GEANT4 simulation of slow positron beam implantation profiles. *Nucl. Instrum. Methods Phys. Res. B* **266**, 4000–4009 (2008).

# Supplementary Information

**Table of contents**

**Supplementary Section 1, Supplementary Fig. 1 (Fig. S1): Fabrication of the as-prepared heterostructure and Supplementary Fig. 2 (Fig. S2): Voltage actuation through electrolyte gating.**

The building block of the heterostructures comprises 45 nm-thick Fe-C films grown at room temperature by co-sputtering onto Si(001) wafers, previously coated with 15 nm Ti and 50 nm Cu. A 5 nm-thick Ti-C film was finally co-sputtered to act as both protective layer from oxidation and carbon ion reservoir. A $Fe_{45}C_{55}$ (in atomic %) film was targeted as a balance between achieving ferromagnetism and facilitating iron carbide formation, aimed at developing a ferromagnetic, magneto-ionic system. Films with low-Fe content lack sufficient Fe percolation[1], preventing ferromagnetism, while those with high-Fe content impede iron carbide formation. Furthermore, high-Fe content films would primarily exhibit electrostatic rather than electrochemical effects[2], due to their large electric conductivity. Fig. S1a shows a representative cross-section transmission electron microscopy (TEM) image of the as-grown heterostructure, evidencing a granular-like Fe-C microstructure[3]. The composition (in atomic %) of the Fe-C and Ti-C layers, quantified using energy dispersive X-ray (EDX) analysis (not shown) over areas of 400 nm$^2$, are $Fe_{45}C_{55}$ and $Ti_{70}C_{30}$, respectively (Methods for considerations on the composition quantification). Fig. S1b is a high-resolution TEM image of the $Fe_{45}C_{55}$ layer, together with the fast Fourier transform (FFT) of the region marked with a yellow square. No spots or rings are visible; instead, a diffuse halo is observed, indicating a lack of crystallinity. This amorphous-like structure is consistent with a Fe-rich phase embedded in a C-rich matrix as previously reported[4,5]. Fig. S1c displays a high-angle annular dark-field scanning transmission electron microscopy (HAADF-STEM) image, together with the C, Fe, Ti, and Cu elemental electron energy loss spectroscopy (EELS) mappings. Both C and Fe are uniformly distributed. As seen in Fig. S1d, the as-grown heterostructure exhibits a clear hysteresis loop, measured by vibrating sample magnetometry (VSM), with a $M_S$ of around 671 emu cm$^{-3}$. This value corresponds to 39 % of the $M_S$ of pure body-centered cubic α-Fe (about 1710 emu cm$^{-3}$), and it is below than the Fe content in the Fe-C layer (45 %), in agreement with a Fe-rich ferromagnetic phase. Furthermore, the hysteresis loop exhibits a squareness $\frac{M_R}{M_S}$ of around 78 % (where $M_R$ is the remanent magnetization). This suggests that the ferromagnetic phase exhibits magnetic anisotropy and has a volume exceeding the superparamagnetic limit. That is, the ferromagnetic constituent is not composed of small, isolated particles but probably a percolated[1] or highly interacting phase[6], in agreement with the compositional analysis through EELS (Fig. S1c).

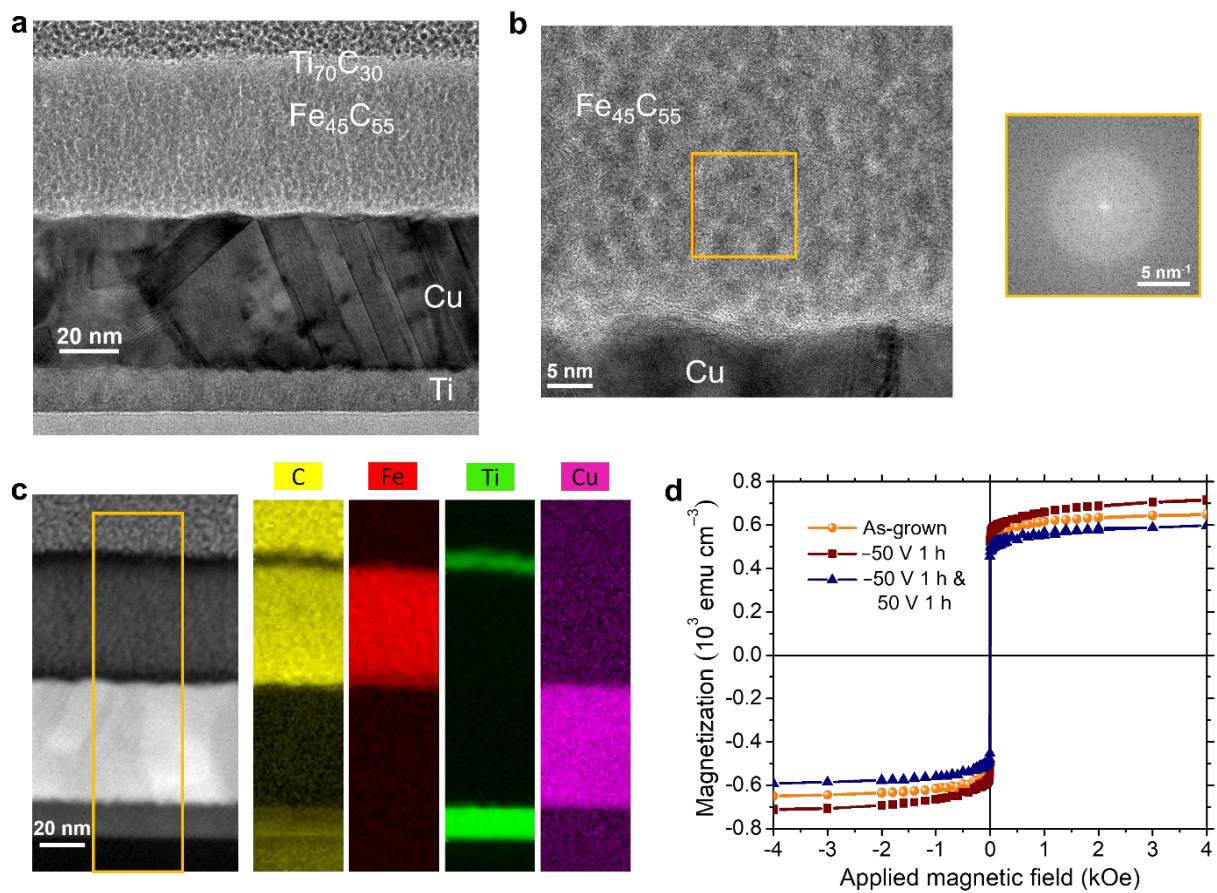

**Fig. S1 | Structural and magneto-ionic characterization of as-grown Fe$_{45}$C$_{55}$-based heterostructures. a**, TEM image of a cross-section of the as-grown heterostructure. **b**, High-resolution TEM image of the Fe$_{55}$C$_{45}$ layer with the FFT of the region enclosed in the yellow square. **c**, HAADF-STEM micrograph and C, Fe, Ti, and Cu elemental EELS mappings of the region marked with a yellow rectangle. **d**, VSM measurements of an as-grown sample, and as-grown samples upon being actuated with −50 V for 1 h, and −50 V for 1 h & 50 V for 1 h. The presence of C on top of the Ti$_{70}$C$_{30}$ layer stems from the Pt-C layer deposited on the samples for the preparation of the TEM lamella (see Methods for further details).

To investigate magneto-electric effects, magnetometry measurements were conducted while electrolyte gating the heterostructure, following the procedures described in Methods and Fig. S2. After biasing at −50 V for 1 h, a slight increase in $M_S$ is observed in the hysteresis loop (Fig. 1d). Subsequently, the system was biased at 50 V for 1 h; the resulting hysteresis loop shows a decrease in $M_S$, reaching a value slightly smaller than that of the as-grown heterostructure (Fig. 1d). This behavior aligns with mild voltage-driven composition changes, likely linked to limited voltage-induced ion transport. Negative biases might slightly lower the C/Fe atomic ratio in the Fe-rich phase (*i.e.*, C content reduction), resulting in a modest $M_S$ increase. In contrast, positive biases may increase the C content in the Fe-rich phase, reducing $M_S$. These restricted magneto-electric effects are linked to the high conductivity of the Fe-rich phase, which shortens the applied electric field[7] (Fig. S3).

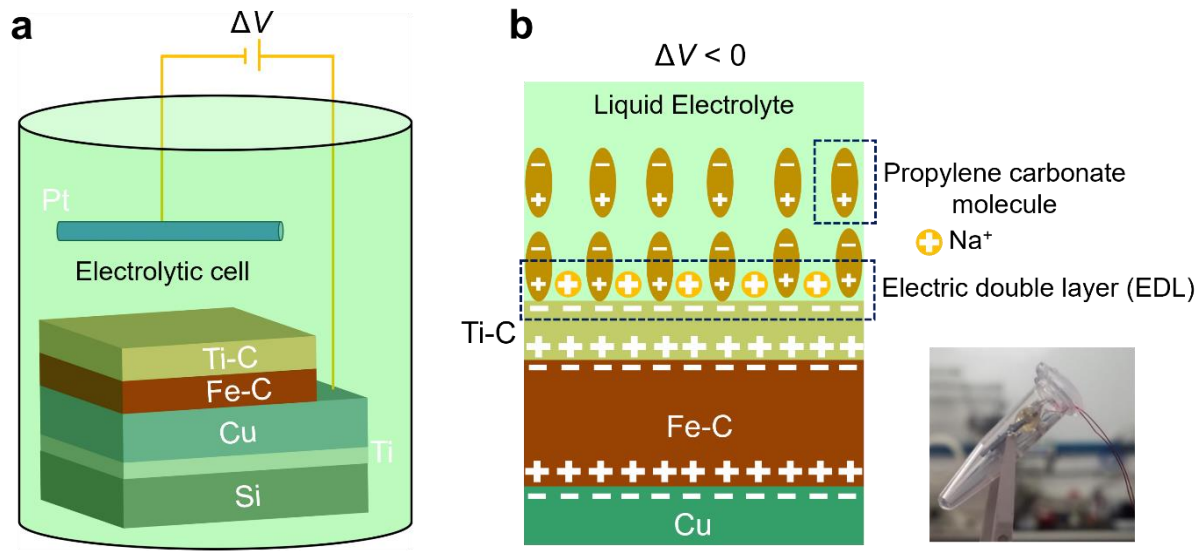

**Fig. S2 | a**, Cartoon of the used electrolytic cell. **b**, Schematic illustration of the electric double layer formation at the interface between the liquid electrolyte and the Ti-C layer when a negative bias is applied to the bottom electrode (*i.e.*, Cu) with respect to the counter electrode (*i.e.*, Pt). On the right, there is a photograph of the custom-built electrolytic cell, built using an Eppendorf and used for voltage actuation.

Subsequently, the heterostructures were partly recrystallized by annealing (Methods)[4] with the goal of inducing the interplay of the following competing processes: (i) Fe and C segregation[1], since carbon has a slightly larger affinity for titanium than for iron[8], and (ii) the formation of non-stoichiometric Fe carbides due to the incomplete recrystallization[4]. (i) would enable a larger electric field penetration[7], as C is more resistive than Fe and the Fe-C system increases resistivity with increased C content —for instance, compared to Fe, resistivity may increase by 25-fold in $Fe_3C$[9] and by a factor over $10^{10}$ in amorphous C[10], whereas (ii) would also increase resistivity while, simultaneously, would favor ion mobility since defects, linked to compositional deficiencies, act as active transport sites in sufficiently resistive systems from an electric viewpoint[11].

**Supplementary Section 2 and Supplementary Fig. 3 (Fig. S3): Electrical characterization.**

As seen in Fig. S3, the annealing results in an increased resistance across the heterostructure. For a given applied voltage, the ratio of electric resistance between the as-annealed and as-grown heterostructures can be determined by dividing the current intensity $I$ of the as-grown sample by that of the annealed one. For instance, for an applied voltage of 0.8 V, $I_{\text{as-grown}} = 1.71 \times 10^{-3}$ A and $I_{\text{annealed}} = 2.75 \times 10^{-7}$ A, the ratio of electric resistance between the as-annealed and as-grown heterostructure is ≈ 6×10$^3$ (~ 10$^4$). Taking into account that, compared to Fe, resistivity may increase by 25-fold in Fe$_3$C[9] and be a factor even larger than 10$^{10}$ in amorphous C[10], thus the increase in resistivity upon annealing is consistent with the formation of amorphous-like C in agreement with the high-resolution TEM analysis of the C-rich areas (*i.e.*, Area I, Area II, and Area III) of the as-prepared sample (Fig. S6a,b). Nevertheless, the following two sources of electrical resistance might also play a role in the observed increase in resistance:

(i) Area I, Area II, and Area III are C-rich regions with compositions in atomic % of Fe$_{15}$C$_{85}$, Fe$_{35}$C$_{65}$, and Fe$_{10}$C$_{90}$, respectively, while Area IV is richer in Fe (Fe$_{65}$C$_{35}$). The layered structure of the sample together with the difference in electrical resistivity between the C-rich region and the Fe-rich layer and between C-rich layers might result in an overall enhanced out-of-plane resistivity due to the formation of Schottky barriers at the interfaces between layers. The top interface between the Ti-C layer and Area I and the bottom interface between Area IV and Cu might also contribute[12].

(ii) The role of surface oxidation of the Ti-C layer by natural passivation in the as-grown sample which might be further increased while annealing due to the residual oxygen traces in the chamber due to the higher affinity of Ti for O than for C[13,14]. Ti oxides usually show high resistivities[15], as large as ~ 10$^7$ μΩ cm (a factor ~ 10$^6$ with respect to the resistivity of Fe).

The presence of residual oxygen throughout the heterostructure is likely possible, such as the oxygen from the naturally passivated top layer of the Cu/Ti/Si substrate. However, the presence of significantly oxidized regions can be ruled out. A high oxygen concentration would compel carbon to adopt a positive oxidation state due to oxygen's much greater electronegativity. This, in turn, would drive carbon ions downward according to our voltage actuation protocol. Since this behavior is not observed, the presence of oxides in the inner regions of the heterostructure can be disregarded.

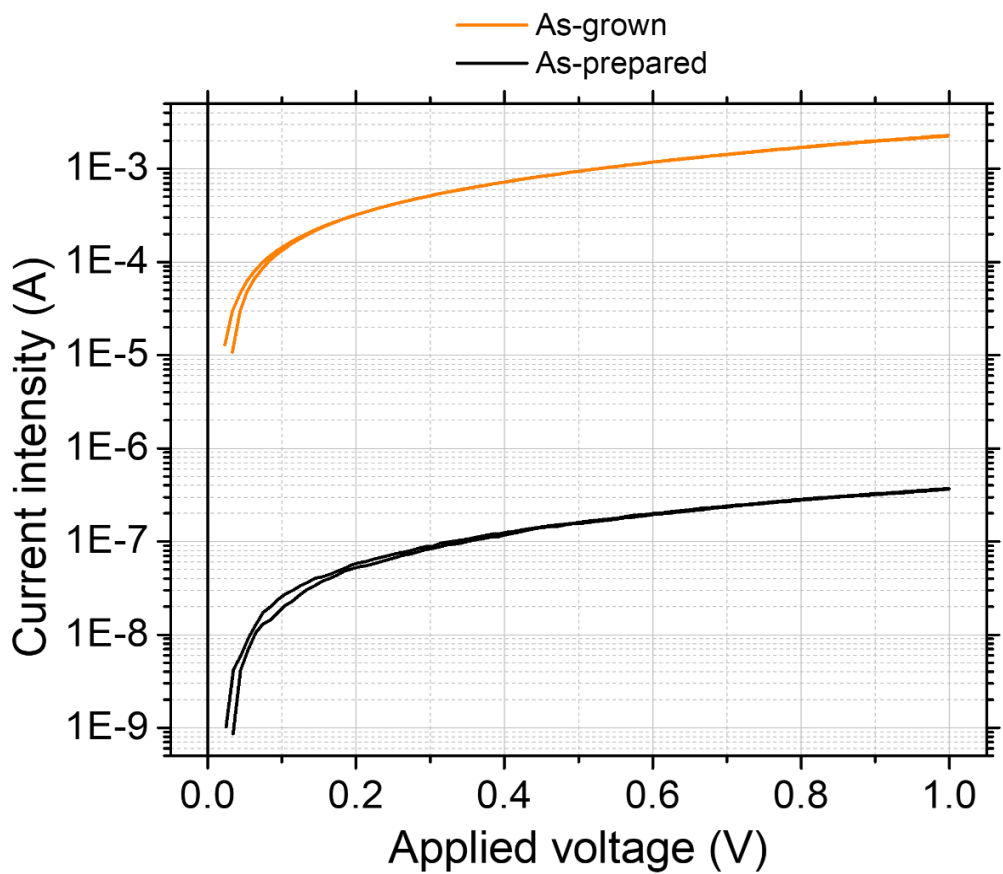

**Fig. S3 |** Current intensity *vs.* applied voltage curves between the Cu bottom electrode and a probe tip of an as-grown and an as-prepared heterostructure.

**Supplementary Fig. 4 (Fig. S4): TEM characterization of the cross-sections of as-grown, as-prepared, and voltage-treated heterostructures.**

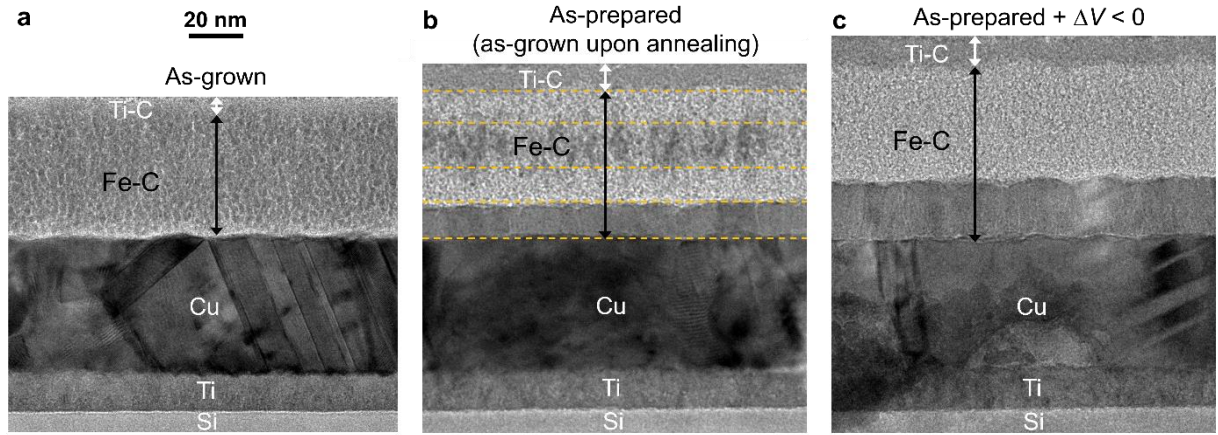

**Fig. S4 |** TEM micrographs of the cross-sections of **a** as-grown and **b** as-prepared heterostructures, and **c** an as-prepared sample subjected to –50 V for 1 h. In panel b, the four-layer structure of the Fe-C film is indicated by yellow dashed lines.

**Supplementary Section 3: Origin of the increase in thickness upon annealing of the as-grown sample (*i.e.*, in the as-prepared heterostructure) and after voltage actuation.**

As shown in Fig. S4, the thickness of the heterostructure increases upon annealing and expands further after voltage actuation. Similar behavior has been observed in other systems during crystallization, as structurally loose interfaces develop[16].

**Supplementary Fig. 5 (Fig. S5): Application of the Avrami formalism to simulate the evolution of the generated magnetic moment at saturation in early stages of voltage actuation.**

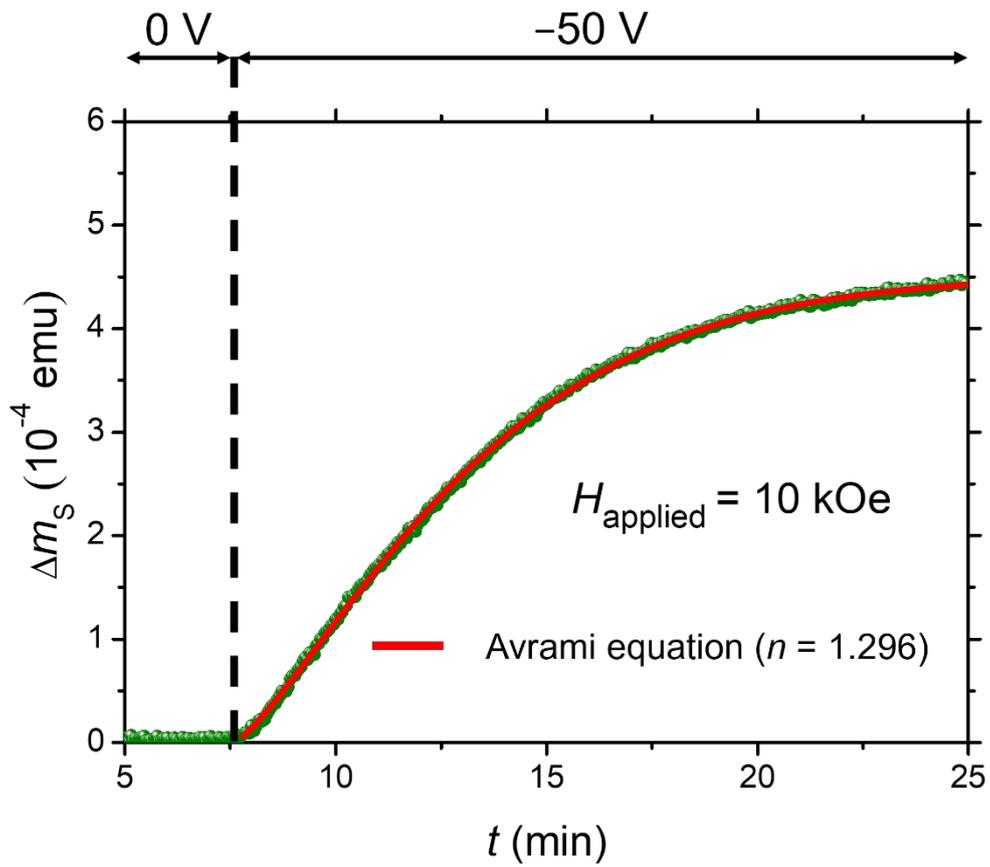

**Fig. S5 |** Dependence of the generated magnetic moment at saturation (*i.e.*, $\Delta m_S$) as a function of time ($t$) while applying –50 V. The experimental datapoints are fitted to an Avrami equation.

**Supplementary Fig. 6 (Fig. S6): High-resolution TEM characterization of Area I, Area II, and Area III of an as-prepared heterostructure, and of Area V of an electrically-treated heterostructure.**

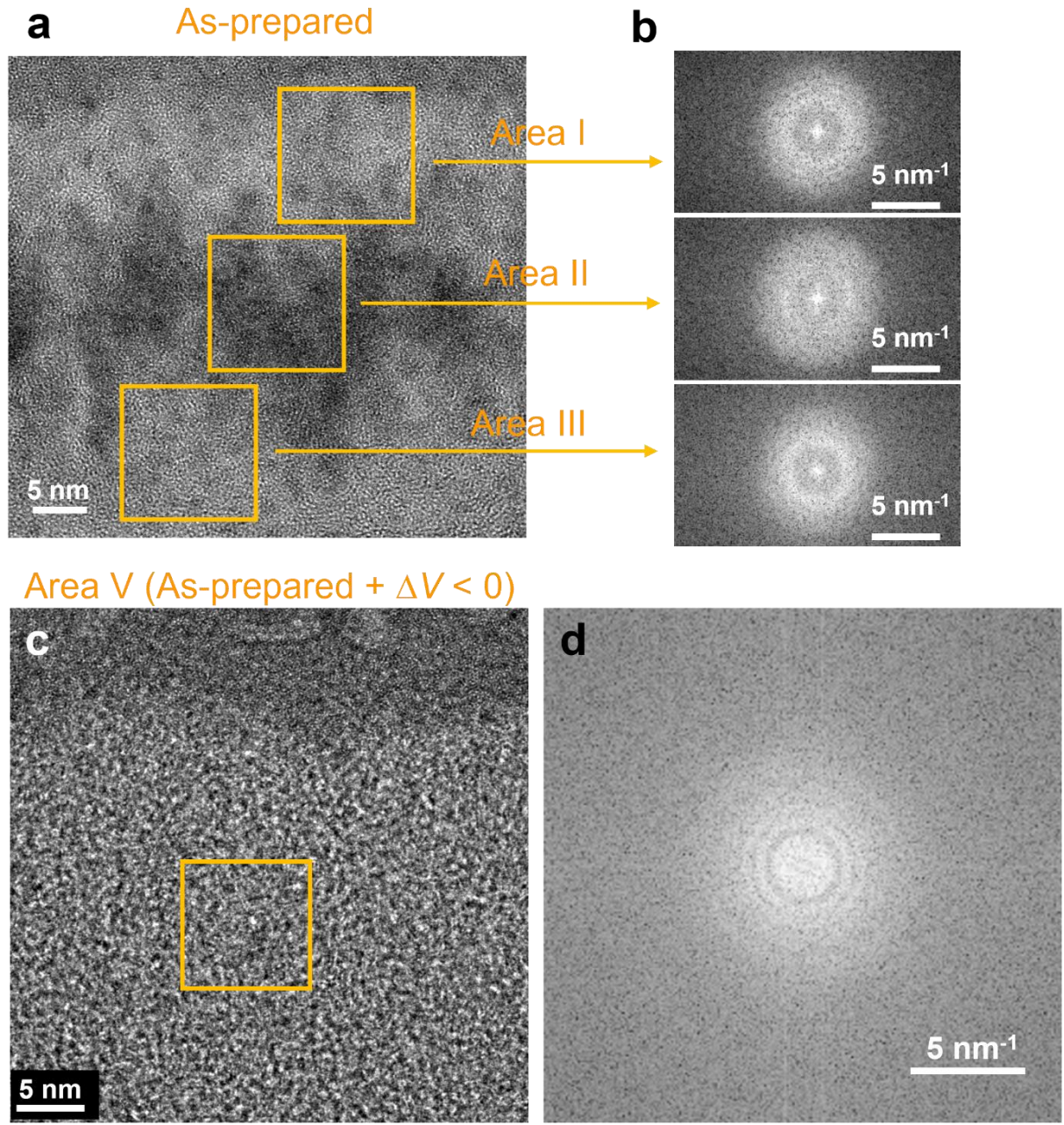

**Fig. S6 | a**, High-resolution TEM micrograph of Area I, Area II and Area III of an as-prepared heterostructure. **b**, FFTs of the regions enclosed in the yellow squares of panel a. **c**, High-resolution TEM micrograph of Area V corresponding to a voltage-treated sample. **d**, FFT of the region enclosed in the yellow square of panel c.

**Supplementary Fig. 7 (Fig. S7): High-resolution TEM characterization of Area IV of an as-prepared heterostructure, and of Area VI of an electrically-treated heterostructure.**

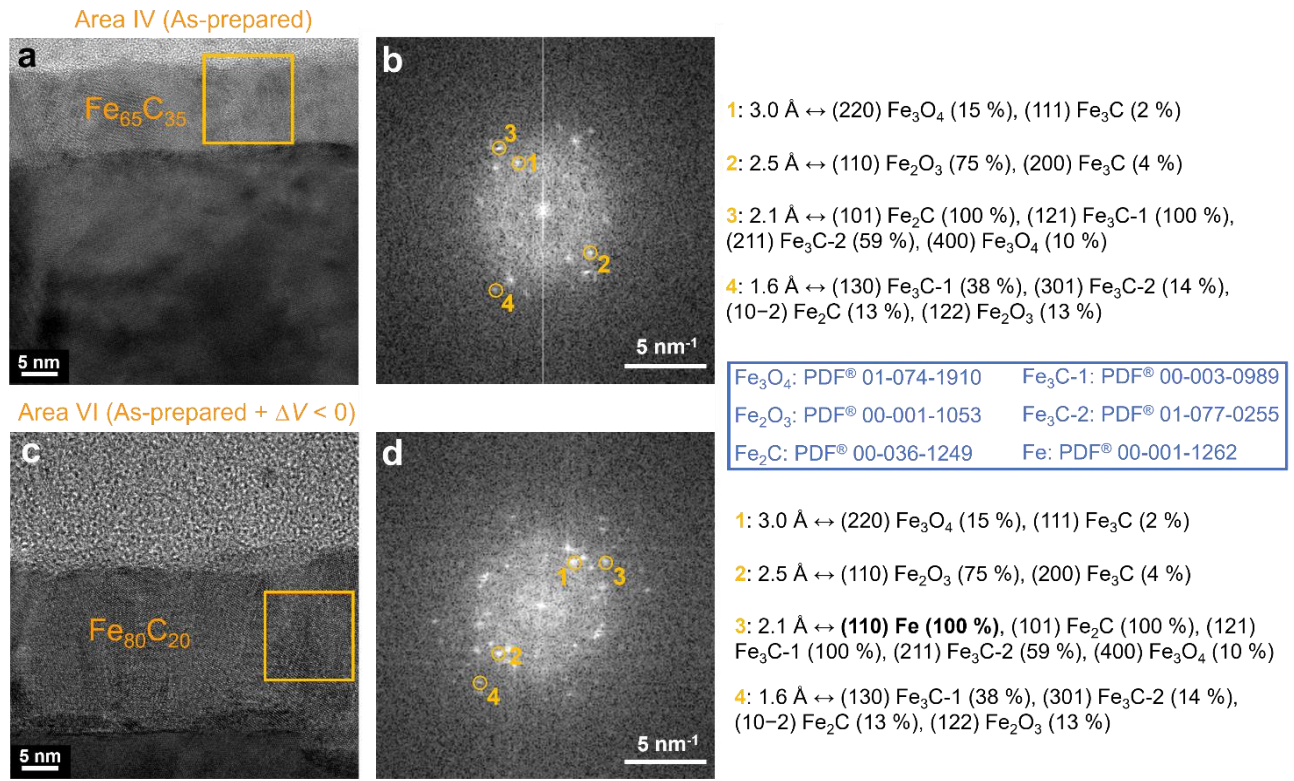

**Fig. S7 | a**, High-resolution TEM micrograph of Area IV corresponding to an-prepared sample. **b**, FFT of the yellow square marked in panel a. **| c**, High-resolution TEM micrograph of Area VI corresponding to a voltage-treated sample. **d**, FFT of the yellow square marked in panel c. On the right side, the diffraction spots are correlated to the phases which they could be compatible with.

**Supplementary Section 4: Further interpretation of the VEPALS characterization.**

The average positron lifetime ($\tau_{Av} = \sum \tau_i I_i$, where i = 1,2), which accounts for the size variations of defects, reduces in the bottom region upon voltage treatment (*i.e.*, Area III and Area IV of the as-prepared sample and Area VI of the biased heterostructure, Fig. 3a). From the spectra decomposition, two major lifetime components (*i.e.*, $\tau_1$ and $\tau_2$) have been extracted and plotted in Fig. 3b. The shorter lifetime indicates small vacancy-like defects, not larger than bi-vacancies and the longer lifetime represents large vacancy clusters of 12 - 16 point defect ($V_{Fe}$) agglomerates as taken from pure, defected iron is given[17,18]. This is also representative of $Fe_3C$ since the calculated lifetime for $Fe_3C$ is only slightly smaller than the one for Fe[19]. As happens with the average positron lifetime, both the shorter lifetime (*i.e.*, $\tau_1$) and the longer lifetime (*i.e.*, $\tau_2$) components decrease upon biasing in the bottom part of the film, indicating a decrease in the vacancy cluster size after voltage actuation. The shorter lifetime contributes to about 45 - 70 % of the annihilation signals ($I_1$ in Fig. 3c) and the longer lifetime counterparts to most of the remaining signals ($I_2$, not shown). The second lifetime component is likely related to defect complexes in the amorphous-like phase of the films, which decreases after biasing. The relative intensities do not change upon treatments, most likely because of the exceedingly large defect concentration (> $10^{19}$ cm$^{-3}$: positron saturation trapping regime).

**Supplementary Fig. 8 (Fig. S8): Compositional analysis through XAS.**

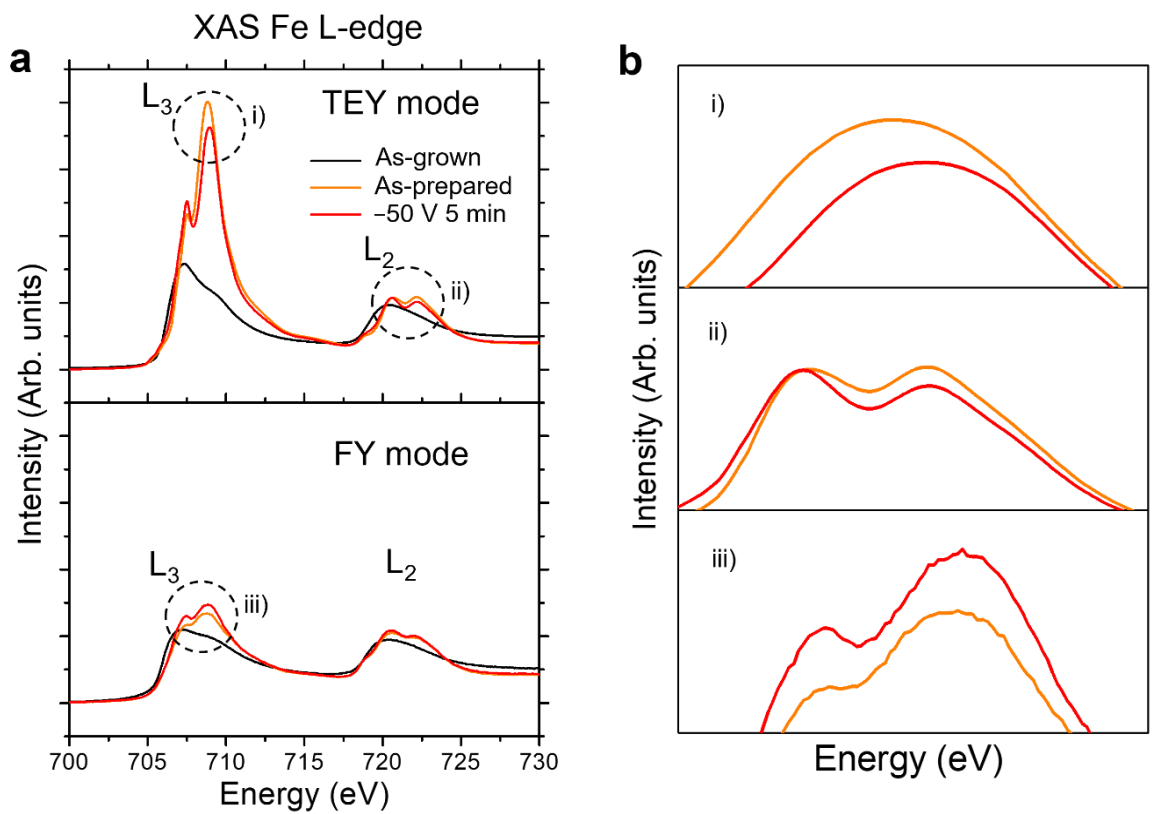

**Fig. S8 | Compositional characterization of early magneto-ionic stages by XAS. a**, Total electron yield (TEY; top) and fluorescence yield (FY; bottom) Fe $L_{2,3}$-edge XAS spectra for an as-grown heterostructure, an as-prepared sample, and a voltage-treated with –50 V for 5 min sample. **b**, Enlargement of the $L_3$ peak region of the spectra shown in panel **a**.

**Supplementary references**

1. Mi, W. B. *et al*. Facing-target sputtered Fe–C granular films: Structural and magnetic properties. *J. Appl. Phys.* **97**, 043903 (2005).

2. Leighton, C., Birol, T. & Walter, J. What controls electrostatic vs electrochemical response in electrolyte-gated materials? A perspective on critical materials factors. *APL Mater.* **10**, 040901 (2022).

3. Mi, W. B. *et al*. Structure and magnetic properties of N-doped Fe–C granular films. *J. Phys. D: Appl. Phys.* **39**, 911–916 (2006).

4. Bauer-Grosse, E. Thermal stability and crystallization studies of amorphous TM–C films. *Thin Solid Films* **447-448**, 311–315 (2004).

5. Furlan, A., Jansson, U., Lu, J., Hultman, L. & Magnuson, M. Structure and bonding in amorphous iron carbide thin films. *J. Phys.: Condens. Matter.* **27**, 045002 (2015).

6. Spaldin, N. A. Magnetic Materials: Fundamentals and Applications, 2nd ed. (Cambridge University Press, 2010).

7. de Rojas, J. *et al.* Critical role of electrical resistivity in magnetoionics. *Phys. Rev. Appl.* **16**, 034042 (2021).

8. Häglund, J., Grimvall, G., Jarlborg, T. & Fernández Guillermet, A. Band structure and cohesive properties of 3d-transition-metal carbides and nitrides with the NaCl-type structure. *Phys. Rev. B* **43**, 14400–14408 (1991).

9. Bhadeshia, H. K. D. H. Cementite. *Inter. Mater. Rev.* **65**, 1–27 (2020).

10. McKenzie, D. R. Tetrahedral bonding in amorphous carbon. *Rep. Prog. Phys.* **59**, 1611–1664 (1996).

11. Ma, Z. *et al.* Controlling magneto-ionics by defect engineering through light ion implantation. *Adv. Funct. Mater.* **34**, 2312827 (2024).

12. Arras, R., Gosteau, J., Tricot, S., & Schieffer, P. Schottky barrier formation at the Fe/SrTiO$_3$(001) interface: Influence of oxygen vacancies and layer oxidation. *Phys. Rev. B* **102**, 205307 (2020).

13. https://webbook.nist.gov/cgi/cbook.cgi?ID=C1317802&Mask=2&utm

14. https://webbook.nist.gov/cgi/formula?ID=C12070085&Mask=2&utm

15. Breckenridge, R. G. & Hosler, W. R. Electrical properties of titanium dioxide semiconductors. *Phys. Rev.* **91**, 793–802 (1953).

16. Zhao, Y., Zhang, B. & Sato K. Unusual volume change associated with crystallization in Ce-Ga-Cu bulk metallic glass. *Intermetallics* **88**, 1–5 (2017).

17. Čížek, J., Melikhova, O., Barnovská, Z., Procházka, I. & Islamgaliev, R. K. Vacancy clusters in ultra fine grained metals prepared by severe plastic deformation. *J. Phys. Conf. Ser.* **443**, 012008 (2013).

18. Ohkubo, H. *et al*. Positron annihilation study of vacancy-type defects in high-speed deformed Ni, Cu and Fe. *Mater. Sci. Eng. A.* **350**, 95–101 (2003).

19. Puska, M. J., Šob, M., Brauer, G. & Korhonen, T. First-principles calculation of positron lifetimes and affinities in perfect and imperfect transition-metal carbides and nitrides. *Phys. Rev. B.* **49**, 10947–10957 (1994).